\documentclass[12pt,preprint]{aastex}
\shorttitle{SMA Observations of CG\,30}
\shortauthors{Chen et al.}

\begin{document}

\title{SMA CO\,(2--1) Observations of CG\,30: a Protostellar Binary System
with a High-Velocity Quadrupolar Molecular Outflow}

\author{Xuepeng Chen\altaffilmark{1,2}, Tyler L. Bourke\altaffilmark{2},
Ralf Launhardt\altaffilmark{1}, Thomas Henning\altaffilmark{1}}

\altaffiltext{1}{Max Planck Institute for Astronomy, K\"{o}nigstuhl 17,
D-69117 Heidelberg, Germany}
\altaffiltext{2}{Harvard-Smithsonian Center for Astrophysics, 60 Garden Street,
Cambridge, MA\,02138, USA}

%%%%%%%%%%%%%%%%%%%%%%%%%%%%%%%%%%%%%%%%%%%%%%%%%%%%%%%%%%%%%%%%%%%%%%%%%%%%
\begin{abstract}

We present interferometric observations in the $^{12}$CO\,(2--1) line and
at 1.3\,mm dust continuum of the low-mass protostellar binary system in the
cometary globule CG\,30, using the Submillimeter Array. The dust continuum
images resolve two compact sources (CG\,30N and CG\,30S), with a linear
separation of $\sim$\,8700\,AU and total gas masses of $\sim$\,1.4 and
$\sim$\,0.6\,$M_\odot$, respectively. With the CO images, we discover two
high-velocity bipolar molecular outflows, driven by the two sources. The two
outflows are nearly perpendicular to each other, showing a quadrupolar
morphology. The northern bipolar outflow extends along the southeast
(redshifted, with a velocity up to $\sim$\,23\,km\,s$^{-1}$) and northwest
(blueshifted, velocity up to $\sim$\,30\,km\,s$^{-1}$) directions, while the
southern pair has an orientation from southwest (blueshifted, velocity up to
$\sim$\,13\,km\,s$^{-1}$) to northeast (redshifted, velocity up to
$\sim$\,41\,km\,s$^{-1}$). The outflow mass of the northern pair, driven by
the higher mass source CG\,30N, is $\sim$\,9 times larger than that of the
southern pair. The discovery of the quadrupolar molecular outflow in the
CG\,30 protobinary system, as well as the presence of other quadrupolar
outflows associated with binary systems, demonstrate that the disks in
(wide) binary systems are not necessarily co-aligned after fragmentation.

\end{abstract}


%%%%%%%%%%%%%%%%%%%%%%%%%%%%%%%%%%%%%%%%%%%%%%%%%%%%%%%%%%%%%%%%%%%%%%%%%%%%
\section{INTRODUCTION}
%%%%%%%%%%%%%%%%%%%%%%%%%%%%%%%%%%%%%%%%%%%%%%%%%%%%%%%%%%%%%%%%%%%%%%%%%%%%

Binarity/multiplicity is frequent among protostellar systems (see e.g.,
Looney et al. 2000). To probe these protostellar binary (protobinary)
systems, high angular resolution observations of gas and optically thin
dust emission at (sub-)\,millimeter (mm) wavelengths are needed, because
protostellar systems are deeply embedded in molecular cloud cores and are
surrounded by infalling envelopes and accretion disks. However, these
observations were long limited by the low angular resolution of single-dish
telescopes and have only become possible with the recent availability of
large (sub-)\,mm interferometers. During the past few years,
interferometric mm observations have enabled us to discover several
protobinary systems (Launhardt 2004), and there is an increasing number
of interferometric studies of binarity in the protostellar stage (e.g.,
Chen et al. 2007). Nevertheless, due to the large effort and observing
time required, the number of known and well-studied protobinary systems
is still very small.

CG\,30 (also known as BHR\,12, Bourke et al. 1995; or DC\,253.3$-$1.6,
Hartley et al. 1986) is a bright-rimmed cometary globule
located in the Gum Nebula region at a distance of $\sim$\,400\,pc
(Reipurth 1983). With high resolution submm (SCUBA; Henning
et al. 2001) and mm (ATCA; Chen et al. 2008, hereafter Paper\,I) dust
continuum observations, we have identified in this globule a wide
protobinary system with a projected separation of $\sim$\,21\farcs7
($\sim$\,8700\,AU) and bolometric luminosities of 13.6\,$\pm$\,0.8 and
4.3\,$\pm$\,0.5\,$L_\odot$ for the northern and southern protostars,
respectively. The two protostars appear to drive their own bipolar
protostellar jets, which are roughly perpendicular to each other, as seen
in the H$_{2}$\,1--0\,$S$(1) (Hodapp \& Ladd 1995) and the $Spitzer$ IRAC
4.5\,$\mu$m images (Paper\,I). However, information on the molecular
outflows in this protobinary system has been limited, although single-dish
observations in the CO line (Nielsen et al. 1998) and CS line (Garay et al.
2002) had shown a large-scale (unresolved) bipolar outflow in the east-west
direction.
In this Letter, we report Submillimeter Array\footnote{The Submillimeter
Array is a joint project between the Smithsonian Astrophysical Observatory
and the Academia Sinica Institute of Astronomy and Astrophysics and is
funded by the Smithsonian Institution and the Academia Sinica.}
(SMA; Ho et al. 2004) observations in the $^{12}$CO\,(2--1) line towards
CG\,30 with angular resolution sufficient to allow us, for the first time,
to map the distribution of outflowing molecular gas from each member of
the protobinary system.

%%%%%%%%%%%%%%%%%%%%%%%%%%%%%%%%%%%%%%%%%%%%%%%%%%%%%%%%%%%%%%%%%%%%%%%%%%%%
\section{OBSERVATIONS AND DATA REDUCTION}
%%%%%%%%%%%%%%%%%%%%%%%%%%%%%%%%%%%%%%%%%%%%%%%%%%%%%%%%%%%%%%%%%%%%%%%%%%%%

Observations at the SMA were carried out on 2008 March 7th in the compact
configuration. Eight antennas were used in the array, providing baselines
with projected lengths from 14.1 to 136\,m. The digital correlator was set
up to cover the frequency ranges 219.5$-$221.3 GHz and 229.5$-$231.3 GHz
in the lower and upper sidebands, respectively. This setup includes three
isotopic CO lines of $^{12}$CO\,(2--1) (230.538\,GHz), $^{13}$CO\,(2--1)
(220.399\,GHz), and C$^{18}$O\,(2--1) (219.560\,GHz), which were observed
with an uniform channel spacing equivalent to 0.5\,km\,s$^{-1}$.
The 1.3\,mm dust continuum emission was also recorded with a total bandwidth
of $\sim$\,3.5\,GHz ($\sim$\,1.7\,GHz USB and $\sim$\,1.8\,GHz LSB).
System temperatures ranged from 200 to 400\,K (depending on elevation),
with a typical value of $\sim$\,300\,K. The primary beam is about 55$''$ at
230\,GHz. The quasar 3C84 was used for bandpass calibration, and the quasars
0747--331 and 0727--115 for gain calibration. Titan was used for absolute
flux calibration, which we judge accurate to $\sim$\,20\%, by comparison
of the final quasar fluxes with the SMA calibration database.

The data were calibrated using the IDL MIR package (Qi 2005) and imaged
using the Miriad toolbox (Sault et al. 1995). With robust {\it uv} weighting
2, cleaned maps of the $^{12}$CO\,(2--1) emission and the continuum emission
were produced with effective synthesized beam sizes of
5\farcs5\,$\times$\,2\farcs4 and 5\farcs0\,$\times$\,2\farcs0,
respectively. The 1\,$\sigma$ rms noise levels are $\sim$\,0.13 (width
$\sim$\,0.5\,km\,s$^{-1}$) and $\sim$\,0.10\,Jy\,beam$^{-1}$ (width
$\sim$\,2.0\,km\,s$^{-1}$) per velocity channel, and
$\sim$\,4\,mJy\,beam$^{-1}$ for the continuum emission. The final
images were uncorrected for the primary beam attenuation. Further
analysis was performed with the
GILDAS\footnote{see http://www.iram.fr/IRAMFR/GILDAS} software package.

%%%%%%%%%%%%%%%%%%%%%%%%%%%%%%%%%%%%%%%%%%%%%%%%%%%%%%%%%%%%%%%%%%%%%%%%%%%%
\section{RESULTS}
%%%%%%%%%%%%%%%%%%%%%%%%%%%%%%%%%%%%%%%%%%%%%%%%%%%%%%%%%%%%%%%%%%%%%%%%%%%%
%%%%%%%%%%%%%%%%%%%%%%%%%%%%%%%%%%%%%%%%%%%%%%%%%%%%%%%%%%%%%%%%%%%%%%%%%%%%
%\subsection{Dust Continuum Emission}
%%%%%%%%%%%%%%%%%%%%%%%%%%%%%%%%%%%%%%%%%%%%%%%%%%%%%%%%%%%%%%%%%%%%%%%%%%%%

The SMA 1.3\,mm dust continuum image of CG\,30 (Fig.\,1) shows two compact
sources, which were previously detected in our ATCA 3\,mm dust continuum
image (see Paper\,I). The angular separation between the two sources is
measured to be 21\farcs8\,$\pm$\,0\farcs5, consistent with the ATCA result.
Following Paper\,I, we refer to the northern source as CG\,30N (also
consistent with IRAS\,08076$-$3556) and to the southern source as CG\,30S.
From Gaussian $uv$ plane fitting, we derive flux densities of
256\,$\pm$\,51\,mJy for source N and of 99\,$\pm$\,20\,mJy for source S. The
large-scale common envelope (see Fig.\,1), detected in the single-dish
1.3\,mm maps with a total flux of $\sim$\,1500\,mJy (R. Launhardt et al.,
in preparation), is resolved out by the interferometer, namely that
$\sim$\,75\% flux is missing in the SMA map.

From the continuum fluxes measured at $\lambda$\,1.3\,mm and 3\,mm,
we derive a spectral index $\alpha$\,$\sim$\,3.2 (where flux
$S_{\nu}$\,$\propto$\,$\nu$$^{\alpha}$) in both sources, suggesting
optically thin emission. The total gas mass in the circumstellar envelope
was then calculated with the same method described in Launhardt \& Henning
(1997). In the calculations, we adopt an interstellar hydrogen-to-dust
mass ratio of 110 (Draine \& Lee 1984) and a factor of 1.36 accounting
for helium and heavier elements. For the two sources, we use a mass-averaged
dust temperature\footnote{The dust temperature has been calculated in
Paper\,I through the results of the spectral energy distribution (SED) and
found to be 22\,K for CG\,30N and 27\,K for CG\,30S. Nevertheless, simple
multi-component greybody fits to an SED to not properly treat temperature
gradients and effects of very small, externally heated grains. Due to this
uncertainty and to make the results for the two sources better comparable,
we use one reference temperature for both sources. The actual dust masses
must be scaled with the (not well-known) dust temperature.} of
$T_{\rm d}$ = 20\,K and an opacity of $\kappa_{\rm 1.3mm}$ =
0.8\,cm$^2$\,g$^{-1}$, a typical value suggested by Ossenkopf \& Henning
(1994) for coagulated grains in protostellar cores. From the 1.3\,mm dust
continuum image, the total gas masses of sources N and S are estimated
to be 1.4\,$\pm$\,0.3\,$M_\odot$ and 0.6\,$\pm$\,0.1\,$M_\odot$,
respectively, for the assumed distance of 400\,pc. By contrast with the
gas masses derived from the 3\,mm dust continuum images, we estimate an
opacity index of $\beta$\,$\sim$\,1.5 ($\kappa_{\nu} \propto \nu^{\beta}$)
for the two sources, a fairly typical index found in protostellar
cores (see Ossenkopf \& Henning 1994).

%%%%%%%%%%%%%%%%%%%%%%%%%%%%%%%%%%%%%%%%%%%%%%%%%%%%%%%%%%%%%%%%%%%%%%%%%%%%
%\subsection{CO\,(2--1) Emission}
%%%%%%%%%%%%%%%%%%%%%%%%%%%%%%%%%%%%%%%%%%%%%%%%%%%%%%%%%%%%%%%%%%%%%%%%%%%%

Figure\,2 shows the velocity channel maps of the $^{12}$CO\,(2--1)
emission. For display purpose, we have smoothed the channel maps into a
velocity resolution of 2.0\,km\,s$^{-1}$. The CO emission is detected
from $V_{\rm LSR}$\,=\,$-$26\,km\,s$^{-1}$ to 48\,km\,s$^{-1}$, with
systemic velocity being $\sim$\,6.5\,km\,s$^{-1}$. We note that the
emission near the systemic velocity is scattered over the field of view,
because the extended emission has been resolved out by the interferometer.
Also note that the emission in the northeast and southeast corners
of the panels between 8 and 18\,km\,s$^{-1}$ is aliasing due to
insufficient $uv$ coverage.
In each panel, the two crosses indicate the positions of the two dust
continuum sources. The CO emission from the two sources can be clearly
distinguished due to their wide separation. For source CG\,30N,
blueshifted emission (from $V_{\rm LSR}$\,=\,$\sim$\,$-$24\,km\,s$^{-1}$
to $\sim$\,3\,km\,s$^{-1}$) is extending to northwest, while redshifted
emission (from $\sim$\,8\,km\,s$^{-1}$ to $\sim$\,29\,km\,s$^{-1}$) is to
southeast; For source CG\,30S, blueshifted emission
(from $\sim$\,$-$6\,km\,s$^{-1}$ to $\sim$\,5\,km\,s$^{-1}$) is extending
to southwest, while redshifted emission (from $\sim$\,30\,km\,s$^{-1}$
is to $\sim$\,48\,km\,s$^{-1}$) to northeast.

Figure\,3 shows the integrated intensity maps of the CO\,(2--1) emission,
plotted on the $Spitzer$ IRAC band\,2 (4.5\,$\mu$m) image. Two separated
bipolar outflows, with projected lengths of $\sim$\,27,000\,AU (northern
pair) and $\sim$\,20,000\,AU (southern pair), are clearly seen in the
images. The position angles (P.A., measured with respect to the red lobe)
are $\sim$\,128$\pm$5$^\circ$ for the northern pair and
$\sim$\,57$\pm$3$^\circ$ for the southern pair, i.e., the two bipolar
outflows are nearly perpendicular to each other. In the northern bipolar
outflow, the blueshifted lobe is leaf-shaped and spatially
coincident with the Herbig-Haro (HH) object HH\,120 (see Fig\,3b), while
the redshifted lobe is double-peaked and pointing to the H$_2$ knot
No.\,6 (Hodapp \& Ladd 1995). H$_2$ knot No.\,2, located to the northwest
of the blueshifted lobe, is also likely to be associated with the CG\,30N
outflow. Centered at CG\,30S is a highly collimated bipolar jet seen in
the IRAC 4.5\,$\mu$m image. The bipolar CO outflow driven by CG\,30S is
spatially coincident with this jet and also shows a collimated morphology,
in particular the redshifted lobe (length to width ratio is $\sim$\,4).
Its blueshifted lobe is spatially coincident with the near-infrared emission
(knot No.\,3), with the dust continuum source being located at the apex of
both the CO and infrared emissions (Fig.\,3c). This is consistent with our
suggestion in Paper\,I that the infrared emission at CG\,30S is due to
scattered light in a cavity evacuated by the jet/outflow.

%%%%%%%%%%%%%%%%%%%%%%%%%%%%%%%%%%%%%%%%%%%%%%%%%%%%%%%%%%%%%%%%%%%%%%%%%%%%
\section{DISCUSSION}
%%%%%%%%%%%%%%%%%%%%%%%%%%%%%%%%%%%%%%%%%%%%%%%%%%%%%%%%%%%%%%%%%%%%%%%%%%%%
%%%%%%%%%%%%%%%%%%%%%%%%%%%%%%%%%%%%%%%%%%%%%%%%%%%%%%%%%%%%%%%%%%%%%%%%%%%%
\subsection{Outflow Physical Parameters}
%%%%%%%%%%%%%%%%%%%%%%%%%%%%%%%%%%%%%%%%%%%%%%%%%%%%%%%%%%%%%%%%%%%%%%%%%%%%

Thanks to the versatility of the SMA correlator, the $^{13}$CO\,(2--1)
line was observed simultaneously with the $^{12}$CO\,(2--1) line towards
CG\,30, which allows us to estimate the outflow mass. For source CG\,30N,
the $^{13}$CO\,(2--1) blueshifted emission is detected at velocities from
$\sim$\,$-$2 to $\sim$\,3\,km\,s$^{-1}$, while redshifted emission is
from $\sim$\,8 to $\sim$\,12\,km\,s$^{-1}$; For source CG\,30S, however,
no $^{13}$CO\,(2--1) outflow gas is found, although weak emission is
detected around systemic velocity. The outflow masses are then calculated
following Cabrit \& Bertout (1990): for the southern bipolar outflow,
as well as high-velocity component of the northern pair where
$^{13}$CO\,(2--1) is not detected, the $^{12}$CO\,(2--1) emission is
assumed to be optically thin, and the mass is calculated at each channel
where outflow emission is detected, and then summed; for the low-velocity
component of the northern pair, a correction for line opacity is performed
at velocities where both the $^{12}$CO/$^{13}$CO\,(2--1) emissions are
detected. In the calculations, we assume LTE conditions, an excitation
temperature of 20\,K, an abundance ratio of [H$_2$]/[$^{13}$CO] equal to
8.9\,$\times$\,10$^{5}$ (Cabrit \& Bertout 1992), and an isotopic ratio
of [$^{12}$CO]/[$^{13}$CO] = 71 (Wilson \& Rood 1994). No correction for
inclination effects was applied in the calculations.

The derived outflow masses, as well as other outflow properties (i.e.,
momentum $P$, energy $E$, mass-loss rate $\dot{M}$$_{\rm out}$, force
$F_{\rm m}$, and mechanical luminosity $L_{\rm m}$), are listed in
Table~2. A simple comparison between the two bipolar outflows shows that
the northern pair, driven by the higher (circumstellar envelope) mass
source CG\,30N, is much stronger than the southern pair ($\sim$\,9
times in mass and $\sim$\,3 times in energy). The outflow parameters
derived in CG\,30 are relatively small in comparison with the general
range of such parameters for sources of similar mass (see Wu et al. 2004).
Nevertheless, we want to mention that the parameters in Table~2 only refer
to the compact outflows detected in the SMA maps. The nature of the
``extended outflow" components is unclear yet and requires confirmation by
combining the SMA data with single-dish data to recover extended missing
flux. Therefore, these parameters represent lower limits. Furthermore, as
suggested by the two larger-scale jets, mosaicked CO observations are
probably needed to fully map the two outflows.

%%%%%%%%%%%%%%%%%%%%%%%%%%%%%%%%%%%%%%%%%%%%%%%%%%%%%%%%%%%%%%%%%%%%%%%%%%%%
\subsection{Binarity v.s. Quadrupolar Outflows}
%%%%%%%%%%%%%%%%%%%%%%%%%%%%%%%%%%%%%%%%%%%%%%%%%%%%%%%%%%%%%%%%%%%%%%%%%%%%

The two bipolar outflows found in CG\,30 resemble a combined quadrupolar
morphology. A similar outflow morphology was also found in several other
(low-mass) sources, like e.g., IRAS\,16293--2422 (Mizuno et al. 1990), 
HH\,288 (Gueth et al. 2001), and L\,723 (Lee et al. 2002 and references 
therein). It is important to note that in all these
sources protobinary systems have been discovered by high-resolution
interferometric observations (IRAS\,16293, Mundy et al. 1992; HH\,288,
Gueth et al. 2001; L\,723, Launhardt 2004). This strongly supports the
hypothesis that many quadrupolar outflows\footnote{Quadrupolar outflows
are sources in which four outflow lobes are observed and seem to be driven
by the same protostellar core (seen in current mm single-dish telescopes).}
are actually two independent outflows driven by a binary system (Anglada
et al. 1991), although several other scenarios have been proposed (see Arce
et al. 2007). Furthermore, assuming that binary systems are formed through
the fragmentation of collapsing protostellar cores (see Goodwin et al. 2007),
these quadrupolar outflows demonstrate that the disks in (wide)
binary systems are not necessarily co-aligned after fragmentation.

On the other hand, if binarity among protostars occurs as frequently
as among pre-main sequence stars, one would expect to find many
quadrupolar/multipolar protostellar outflows. Such kind of outflows are
however rare in our observations. One possible reason is that
protostellar disks (jets) are preferentially aligned during
fragmentation, and they may then drive one common large-scale outflow.
An example is the source HH\,30 (see Guilloteau et al. 2008 and references
therein). At the same time, the outflow strength of protostars is
proportional to the circumstellar envelope mass (Bontemps et al. 1996).
Since binary protostars appear to have very unequal circumstellar masses
(Launhardt 2004), their relative outflow strengths are expected to be also
very unequal. Therefore, another possible reason is that the weaker
outflow may often not be detected with current single-dish telescopes
that do not have sufficient angular resolution and sensitivity. The
discovery of an equal-mass protobinary system in L\,723 (Launhardt 2004)
and IRAS\,16293 (Mundy et al. 1992) and of a weak secondary outflow from
the lower-mass component in the unequal-mass protobinary systems CG\,30
(this work), CB\,230 (Launhardt 2004), and BHR\,71 (Bourke 2001, Parise et
al. 2006) supports this explanation. However, high-resolution outflow data
so far exist only for very few systems, and it is unclear yet why multipolar
protostellar outflows are so rare in our observations.

%%%%%%%%%%%%%%%%%%%%%
\acknowledgments
%%%%%%%%%%%%%%%%%%%%%
We thank the SMA staff for technical support during the observations, and
Mark Gurwell for his maintenance of the Submillimeter Calibration List.
Support for this work was provided by NASA through contracts 1224608 and
1279198 (T.~L.~B.) issued by JPL, Caltech, under NASA contract 1407.

%\clearpage
%%%%%%%%%%%%%%%%%%%%%%%%%%%%%%%%REFERENCES%%%%%%%%%%%%%%%%%%%%%%%%%%%%%%%%%%

\clearpage

%%%%%%%%%%%%%%%%%%%%%%%%%%%%%%%%TABLES%%%%%%%%%%%%%%%%%%%%%%%%%%%%%%%%%%%%

%%%%%%%%%%%%%%%%%%%%%%%%%%%%%% SMA 1.3mm %%%%%%%%%%%%%%%%%%%%%%%%%%%%%%%%%
\begin{deluxetable}{lcccc}
\tabletypesize{\scriptsize} \tablecaption{SMA 1.3\,mm dust continuum results of CG\,30\label{tbl-1}} \tablewidth{0pt}
\tablehead{\colhead{Source} &\colhead{$S_{\nu}$$^{a}$} & \multicolumn{2}{c}{FWHM sizes$^{a}$} &\colhead{$M_{\rm g}$$^b$} \\
\cline{3-4}\colhead{}       & \colhead{[mJy]}    & \colhead{maj.$\times$min.}& \colhead{P.A.} & \colhead{[$M_{\odot}$]}}
\startdata
CG\,30N   & 256$\pm$51  & 3\farcs4$\times$1\farcs8 &$-$86$\pm$5\degr  & 1.4$\pm$0.3 \\
CG\,30S   &  99$\pm$20  & 5\farcs5$\times$2\farcs3 &   11$\pm$7\degr  & 0.6$\pm$0.1 \\
\enddata
\tablenotetext{a}{Flux densities and FWHM sizes of the continuum sources derived from Gaussian $uv$ plane fitting.
The error bar of flux density is derived from $\sqrt{\sigma^2_{\rm cali}+\sigma^2_{\rm fit}}$, where $\sigma_{\rm cali}$
is the uncertainty from calibration ($\sim$\,20\% of flux density) and $\sigma_{\rm fit}$ is the uncertainty from Gaussian fitting.}
\tablenotetext{b}{Total gas mass; See text for dust temperature and opacity used.}
\end{deluxetable}
%%%%%%%%%%%%%%%%%%%%%%%%%%%%%%%%%%%%%%%%%%%%%%%%%%%%%%%%%%%%%%%%%%%%%%%%%%

%%%%%%%%%%%%%%%%%%%%%%%%%%%%%% SMA outflow %%%%%%%%%%%%%%%%%%%%%%%%%%%%%%%
\begin{deluxetable}{rcccccc}
\tabletypesize{\scriptsize}\tablecaption{\footnotesize Outflow parameters\label{tbl-1}} \tablewidth{0pt} \tablehead{\colhead{}&\colhead{Mass$^a$}&\colhead{Momentum$^a$}&\colhead{Energy$^a$}&\colhead{$\dot{M}$$_{\rm out}$$^b$}&\colhead{Force$^b$}&\colhead{Luminosity$^b$}\\
\colhead{Component}&\colhead{[10$^{-3}$\,$M_\odot$]}&\colhead{[$M_\odot$\,km\,s$^{-1}$]}&\colhead{[10$^{43}$\,ergs]}&\colhead{[10$^{-6}$\,$M_\odot$\,yr$^{-1}$]}
&\colhead{[10$^{-5}$\,$M_\odot$\,km\,s$^{-1}$\,yr$^{-1}$]}&\colhead{[$L_\odot$]}} \startdata

CG\,30N blue & 17.4 & 0.19  & 3.1 & 7.3 &  8.1 & 0.11\\
        red  & 37.9 & 0.48  & 8.7 & 16  & 19.8 & 0.30\\
CG\,30S blue &  3.6 & 0.047 & 0.6 & 2.1 &  2.7 & 0.03\\
        red  &  2.3 & 0.095 & 4.0 & 1.3 &  5.6 & 0.19\\
\enddata
\tablenotetext{a}{The lower limits derived from the SMA observations.}
\tablenotetext{b}{The outflow mass-loss rate, force, and mechanical
luminosity are estimated from the mass, momentum, and energy
with dynamical ages ($\sim$\,2400\,yr for the northern pair and
$\sim$\,1700\,yr for the southern pair). These dynamical ages only
refer to the compact outflows detected in the SMA maps.}
\end{deluxetable}

%%%%%%%%%%%%%%%%%%%%%%%%%%%%%%%FIGURES%%%%%%%%%%%%%%%%%%%%%%%%%%%%%%%%%%%%
\clearpage

\begin{figure}
\begin{center}
\includegraphics[width=10cm,angle=0]{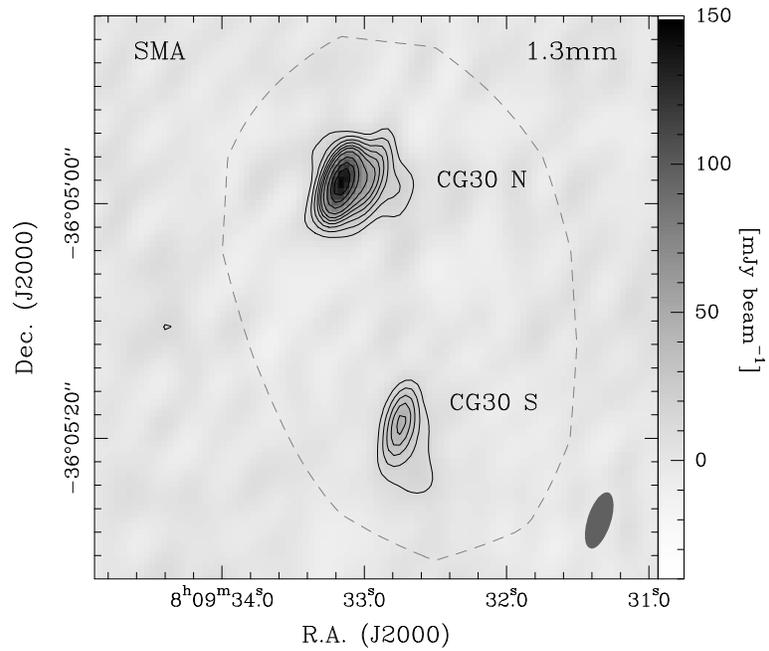}
\caption{SMA 1.3\,mm dust continuum image of CG\,30. Contours start at
$\sim$\,3\,$\sigma$ (1\,$\sigma$\,$\sim$\,4\,mJy\,beam$^{-1}$) with steps
of $\sim$\,2\,$\sigma$ to 17\,$\sigma$, and then increase with steps of
$\sim$\,5\,$\sigma$. The grey dashed contour represents the half-maximum
level of the 1.3\,mm continuum emission observed with SEST (Launhardt et
al. in prep.). The synthesized SMA beam is shown as a grey oval in the
bottom right corner.\label{1mmc}}
\end{center}
\end{figure}

%%%%%%%%%%%%%%%%%%%%%%%%%%%%%%%FIGURES%%%%%%%%%%%%%%%%%%%%%%%%%%%%%%%%%%%%
\begin{figure*}
\begin{center}
\includegraphics[width=16cm, angle=0]{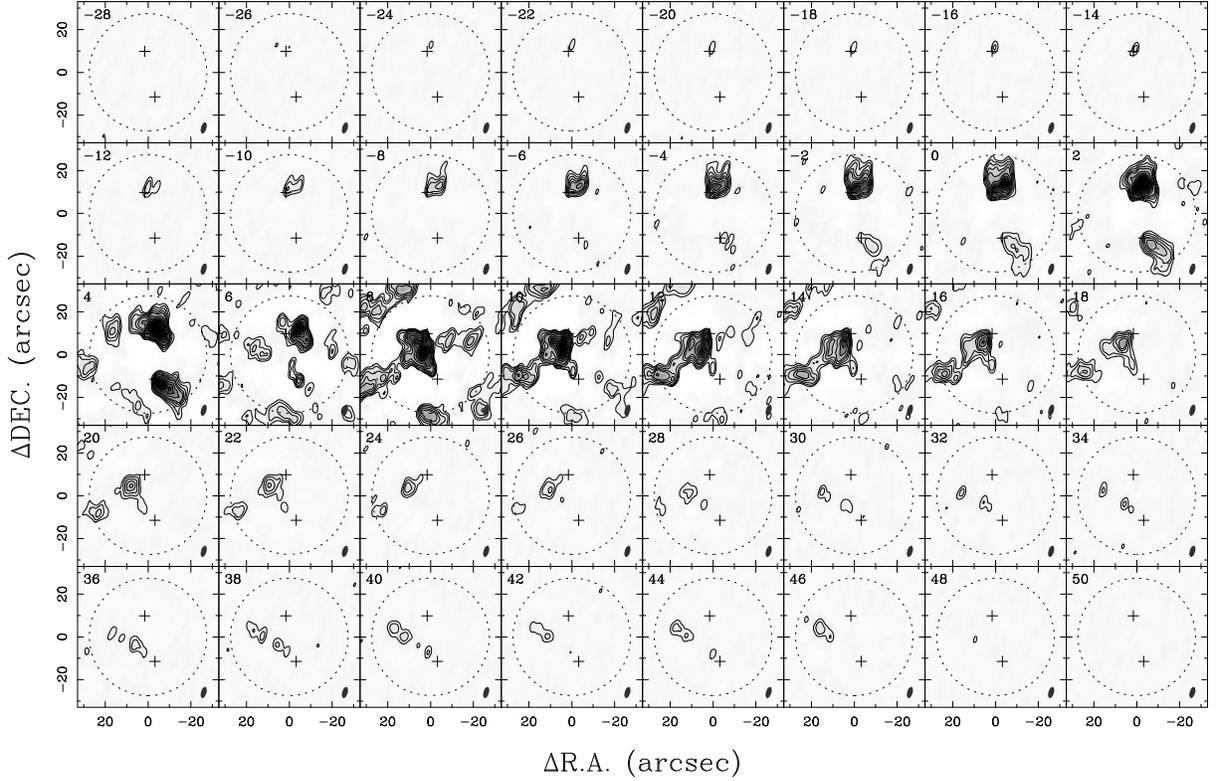}
\caption{Velocity channel maps of the $^{12}$CO\,(2--1) emission
(phase center R.A.=08:09:32.998, DEC=$-$36:05:08.00, J2000). The channels
were smoothed into 2.0\,km\,s$^{-1}$ per step from the original velocity
resolution of 0.5\,km\,s$^{-1}$. The center velocity of each channel is
written in top left corner of each panel (in km\,s$^{-1}$). The systemic
velocity of the molecular cloud is $\sim$\,6.5\,km\,s$^{-1}$. Contours
levels correspond to 3, 6, 10, 15, and 20\,$\sigma$, then increase in
steps of 10\,$\sigma$, where the 1\,$\sigma$ level is
$\sim$\,0.1\,Jy\,beam$^{-1}$. In each panel, the crosses mark the
positions of the two dust continuum sources, the open dotted circle
shows the primary beam ($\sim$\,55$''$ at 230\,GHz), and the filled
ellipse (lower right corner) indicates the synthesized beam
of SMA.\label{channel}}
\end{center}
\end{figure*}

%%%%%%%%%%%%%%%%%%%%%%%%%%%%%%%FIGURES%%%%%%%%%%%%%%%%%%%%%%%%%%%%%%%%%%%%

\begin{figure*}
\begin{center}
\includegraphics[width=16cm, angle=0]{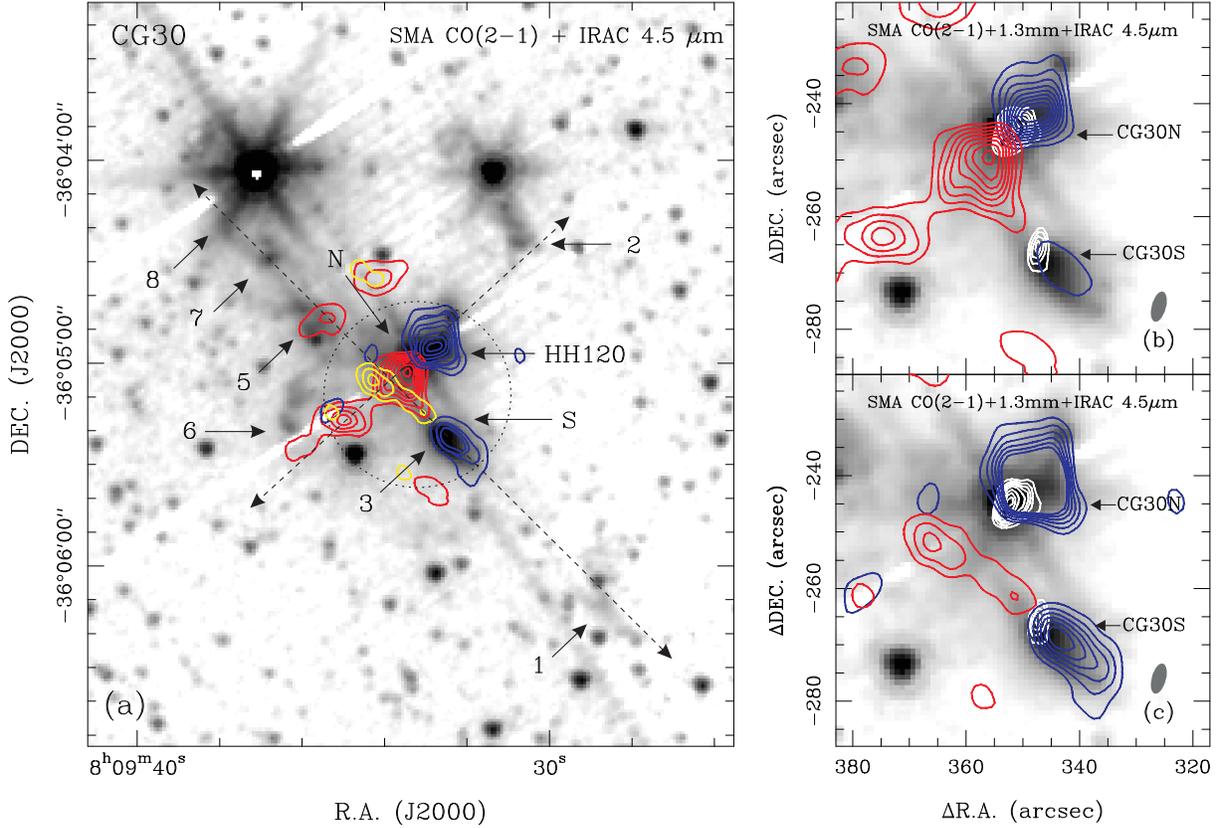}
\caption{(a) The integrated intensity maps of the $^{12}$CO\,(2--1) emission,
plotted on the $Spitzer$ IRAC\,2 (4.5\,$\mu$m) image of CG\,30 (from Paper\,I).
The SMA CO contours (color) are described below. The dotted circle shows the
primary beam of SMA. Sources CG\,30N and CG\,30S are labeled as ``N" and ``S",
respectively. The two dashed arrows show the directions of the
infrared jets, while the arrow delineating the CG\,30N jet, redefined by the
bipolar CO outflow, is $\sim$\,20 degrees different to the one marking the
same jet shown in Fig.\,7 of Paper\,I. The knots in the jets are labeled with
the same numbers as in Hodapp \& Ladd (1995); (b) An enlarged view of the
IRAC\,2 image, overlaid with the SMA 1.3\,mm dust continuum contours (white)
and the $^{12}$CO\,(2--1) emission contours (reference position at
R.A.=08:09:04.082, DEC=$-$36:00:53.53, J2000). The map shows the bipolar
outflow driven by CG\,30N. The blue (red) contours represent emission
integrated over the velocity range
$-$24\,km\,s$^{-1}$\,$<$\,$V_{\rm LSR}$\,$<$\,3\,km\,s$^{-1}$
(8\,km\,s$^{-1}$\,$<$\,$V_{\rm LSR}$\,$<$\,29\,km\,s$^{-1}$), which is
blueshifted (redshifted) respect to the cloud systemic velocity. SMA CO
contours are from 10\,Jy\,km\,s$^{-1}$ by step of 10\,Jy\,km\,s$^{-1}$.
(c) Same as Fig.\,3b, but the map shows the bipolar outflow driven by
CG\,30S. The blue (red) contours represent emission integrated over the
velocity range $-$6.5\,km\,s$^{-1}$\,$<$\,$V_{\rm LSR}$\,$<$\,5\,km\,s$^{-1}$
(30\,km\,s$^{-1}$\,$<$\,$V_{\rm LSR}$\,$<$\,48\,km\,s$^{-1}$).
SMA CO contours are from 5\,Jy\,km\,s$^{-1}$ by step of 5\,Jy\,km\,s$^{-1}$.
For display purpose, we use yellow contours in Fig.\,3a to show the redshifted
emission from source CG\,30S.\label{outflow}}
\end{center}
\end{figure*}
\end{document}